\documentclass[a4,draftcls,12pt,onecolumn]{IEEEtran}
\usepackage{cite}
\usepackage{epsfig,graphics,mathrsfs,amssymb,amsmath,bm,color,yfonts,txfonts,multirow,multicol,slashbox,dblfloatfix}
\usepackage{subfigure}
\usepackage{algorithm2e}
\usepackage[noend]{algpseudocode}

\begin{document}

\title{Resource Allocation in Shared Spectrum
Access Communications for Operators with
Diverse Service Requirements}
\author{Mirza Golam Kibria, \textit{Member, IEEE,} Gabriel Porto Villardi, \textit{Senior Member, IEEE,}  Kentaro Ishizu,  Fumihide Kojima, \textit{Member, IEEE} and Hiroyuki Yano
 \thanks{The authors are with Smart Wireless Laboratory, Wireless Network Research Institute, National Institute of Information and Communications Technology (NICT), 3-4 Hikarino-oka, Yokosuka, Japan 239-0847 (e-mail: mirza.kibria@nict.go.jp, gpvillardi@nict.go.jp, ishidu@nict.go.jp, f-kojima@nict.go.jp, yano@nict.go.jp). }
}
\maketitle

\begin{abstract}

In this paper, we study inter-operator spectrum sharing and intra-operator resource allocation in shared spectrum access communication systems and propose efficient dynamic solutions to address both inter-operator and intra-operator resource allocation optimization problems. For inter-operator spectrum sharing, we present two competent approaches, namely the subcarrier gain based sharing and fragmentation based sharing, which carry out fair and flexible allocation of the available shareable spectrum among the operators subject to certain well-defined sharing rules, traffic demands and channel propagation characteristics. Subcarrier gain based spectrum sharing scheme has been found to be more efficient in terms of achieved throughput. However, fragmentation based sharing is more attractive in terms of computational complexity. For intra-operator resource allocation, we consider resource allocation problem with users' dissimilar service requirements, where the operator supports users with delay-constraint and non-delay constraint service requirements, simultaneously. This optimization problem is a mixed integer nonlinear programming problem and nonconvex, which is computationally very expensive, and the complexity grows exponentially with the number of integer variables. We propose less-complex and efficient suboptimal solution based on formulating exact linearization, linear approximation and convexification techniques for the nonlinear and/or non-convex objective functions and constraints. Extensive simulation performance analysis has been carried out that validates the efficiency of the proposed solution. 
\end{abstract}

\IEEEpeerreviewmaketitle
\section{Introduction}
\label{section:chapter3-1}
Frequency spectrum is an extremely valuable and important natural resource. The exponential increase in demand for the technologies like Wi-Fi or smart electricity grids means we must utilize this finite radio resource very efficiently. But matching this exponentially growing demand for wireless connectivity is harder in the absence of unused or vacant spectrum. In traditional exclusive licensing systems, many frequency bands are spatially and temporally underutilized. Due to the deficiency of the spectrum resources and to support the predicted enormous wireless traffic explosion in future, it is important to make full use of the existing radio resources. Spectrum sharing presents a supplementary approach to conventional license-exempt and exclusive licensing schemes, and can be realized to cope with the existing network infrastructure with the support of new technologies. Even though many applications still depend on exclusive access to spectrum, spectrum sharing \cite{Irnich,Matinmikko,Yrjola} is increasingly recognized as the breeding framework for wireless innovation that triggers the development and deployment of more resilient and flexible wireless technologies.

Spectrum sharing among operators can appear in many different scenarios. One example is co-primary sharing, where the spectrum regulator licenses a frequency band to multiple operators without specifying the boundaries between the bands of spectrum sharing operators and all the operators have equal right to access the shareable spectrum. Another example is licensed shared access scenario, where an incumbent user licenses its frequency band to multiple operators for shared usage in a certain geographical location and for a certain time period. Spectrum sharing is coordinated in accordance with sharing rules under a well-defined set of conditions and mutual agreement. Shared-spectrum access \cite{Mustonen,Jush,Gundlach,Abitbol,Mueck,Palolo} facilitates efficient utilization of the available spectrum in 5th generation (5G) and beyond networks, and will become unquestionably mandatory in order to accommodate the predicted enormous wireless traffic explosion. It acts as an intermediary solution between conventional unlicensed and licensed strategies in which the spectrum sharing operators share the licensed spectrum under a decided set of coverage restrictions and time-period. Furthermore, spectrum sharing represents a supplementary approach to conventional license-exempt and exclusive licensing schemes, and can be realized to cope with the existing network infrastructure with reasonable and feasible modifications \cite{Ahokangas}. 

{\bf{\textit{Related Works and Issues on Inter-operator Spectrum Sharing.}}}
Dynamic and flexible inter-operator spectrum sharing among the participating operators is very important in shared spectrum access scenarios. A large number of issues have to be considered, such as, the spectrum sharing policy, operators' individual traffic demands, structure of the shared-spectrum, i.e., contiguous/non-contiguous, the operating environment, channel propagation characteristics, inter-operator interference, etc. Unlike, resource allocation in other systems, inter-operator resource allocation in shared-spectrum communication depends on various factors such as license agreement policy, traffic demands along with other conventional constraints \cite{Perez,Luo}. 

The work in \cite{Luo} considered a shared spectrum access model, where each operator is allocated with a fragment of the shareable spectrum. However, \cite{Luo} does not consider dynamic fragment sharing among the operators, which can have a significant impact on the system performance. This is because the achievable throughput for a {\bf{spectrum}} sharing operator over different fragments of the shared spectrum can vary quite significantly depending on the types of applications and channel characteristics. In \cite{Anchora}, a centralized approach for spectrum sharing across multiple operators has been proposed based on a coordinated scheduling algorithm. The authors of \cite{Kamal, Bennis} considered {\it{orthogonal frequency division multiple access}} (OFDMA) based scheme and proposed spectrum sharing approaches from a game theoretic perspective under cognitive radio context, where the spectrum sharing operators are classified as primary and secondary. Unfortunately, each of the above mentioned shared spectrum allocation works considered either subcarrier gain based or fragmentation based schemes, not both.

Furthermore, during the inter-operator spectrum sharing process, if an operator needs to ensure that all the users have approximately same data rates or each user should be able to transmit at a minimum rate, some notion of fairness has to be incorporated in the optimization process that gives the users the way of being treated in accordance with the fairness notion. In this study, we aim at treating all the users equally in terms of allocating resources to them. Along with achieving overall higher system throughput, maximizing fairly shared spectrum efficiency is very important, especially from the spectrum sharing operators' perspectives.

{\bf{\textit{Related Works and Issues on Intra-operator Resource Allocation.}}}
From the allocated spectrum in the inter-operator spectrum sharing stage, each operator then allocates the radio resources to its own users depending on system objective, users' applications types and other constraints. In general, the operators can perform such resource allocation independently from each other. For the intra-operator resource allocation, we consider non-overlapping subcarrier allocation and the operators support users with heterogenous service requirements. There are many works in literature dealing with the problem of resource allocation in OFDMA system under various system constraints \cite{Rhee,Jang,Yu,Ng,Wong}. For instance, the authors of \cite{Rhee,Jang} have shown that the overall system capacity of an OFDMA system is optimized when each subcarrier is assigned to the user with the best channel gain. The \textit{max-min} optimization problem is addressed in \cite{Rhee}, where all the users are assured to achieve a similar data rate through the maximization of the worst users' capacity. 

The algorithm proposed in \cite{Jang} is aimed at the maximization of data rate under total transmitting power and target bit error rate requirements. In \cite{Yu,Ng}, the authors claim that non-convexity is not an issue for the resource allocation problem in an OFDMA system if the number of subcarriers is very large. In \cite{Wong}, the authors proposed an iterative resource allocation algorithm to minimize the total transmitting power under fixed user data rates and bit error rate constraints. In \cite{Bae}, the authors proposed a best-effort fairness scheme that ensures minimum number of subchannels for all the users. In \cite{Hu,Xia}, adaptive resource allocation in OFDMA system is considered under partial channel state information. In \cite{Shen,Wong1,Sadr1,Mohanram}, the authors formulate an optimization problem, which balances the trade-off between capacity and fairness among the users. Proportional fairness is assured, i.e., ensures that the rates of different users are proportional, by imposing a set of nonlinear constraints.

We consider intra-operator resource allocation for a system with users with delay constraint service requirements. The users are categorized under two difference sets, i.e., {\it{delay-constraint}} (DC) users and {\it{non-delay-constraint}} (NDC) users. Unlike \cite{Wong} and \cite{Jang,Tsang}, which consider homogenous traffic for DC users and NDC users, respectively, in our considered system model, both DC and NDC traffics can be supported simultaneously. Due to the presence of nonlinear structure in the objective function and constraints, and non-sharing nature of the subcarrier allocation among the users, the optimization problem becomes a {\it{mixed-integer nonlinear programming}} (MINLP) problem, which is computationally very expensive. In \cite{Tao}, the authors studied the resource allocation problem in a system with users requiring delay differentiated services and consider fairness in terms of delay sensitive users. They proposed a suboptimal solution by introducing time-sharing variables, and therefore, the system model employed differs from the original OFDMA system. 

{\bf{\textit{Contributions.}}} 
For inter-operator spectrum sharing in shared spectrum access communications, we propose two solutions that are computationally inexpensive. For both of the solutions, optimizing total system throughput has been the objective metric while allocating spectrum resources to the operators. The first solution (a.k.a. Subcarrier gain based spectrum sharing) is iterative in nature. In this proposed scheme, unlike \cite{Anchora,Kamal,Bennis}, we emphasize on fairness issues not only for the spectrum sharing operators, but also for the users served by the operators, by taking care of the sharing policy measures based on well-defined sharing conditions, traffic demands and propagation environments. The second proposed solution (a.k.a. Fragmentation based spectrum sharing) is based on allocating spectrum fragments to the operators instead of subcarriers as in the case of the first solution, where each fragment is a set of larger number of contiguous subcarriers. Note that each operator can dynamically obtain multiple non-contiguous fragments from the shared spectrum if the available spectrum for sharing is non-contiguous.
Contrary to the shared spectrum scheme proposed in \cite{Anchora}, the proposed solution in this work allows dynamic sharing of frequency spectrum among the operators depending on the types of applications, e.g., short-range communications, the technologies it operates on and channel propagation characteristics.

For the intra-operator resource (spectrum and power) allocation problem, we propose a computationally efficient (if not, at least solvable) solution based on some linearization techniques (exact linearization or linear approximations) considering the structures of the optimization problem and the constraints. The performance of the proposed solution is impressive when compared to the original MINLP and other existing solutions. In particular, we transform the computationally expensive nonconvex MINLP into a convex problem by introducing a series of efficient linearization, linear approximation and convexification techniques, therefore, significantly reduce the computational time.

The reminder of this paper is structured as follows. The system model and the problem statement are discussed in Section 2. In Section 3, we discuss two competent solutions for inter-operator spectrum sharing. In Section 4, the proposed scheme for intra-operator resource allocation for users with dissimilar services is discussed. In Section 5, we describe the simulation parameters and evaluate the performances of the proposed solutions. Finally, we conclude the paper in Section 6.

\section{System Model }
\begin{figure}
  \centering
   \includegraphics[scale=.07]{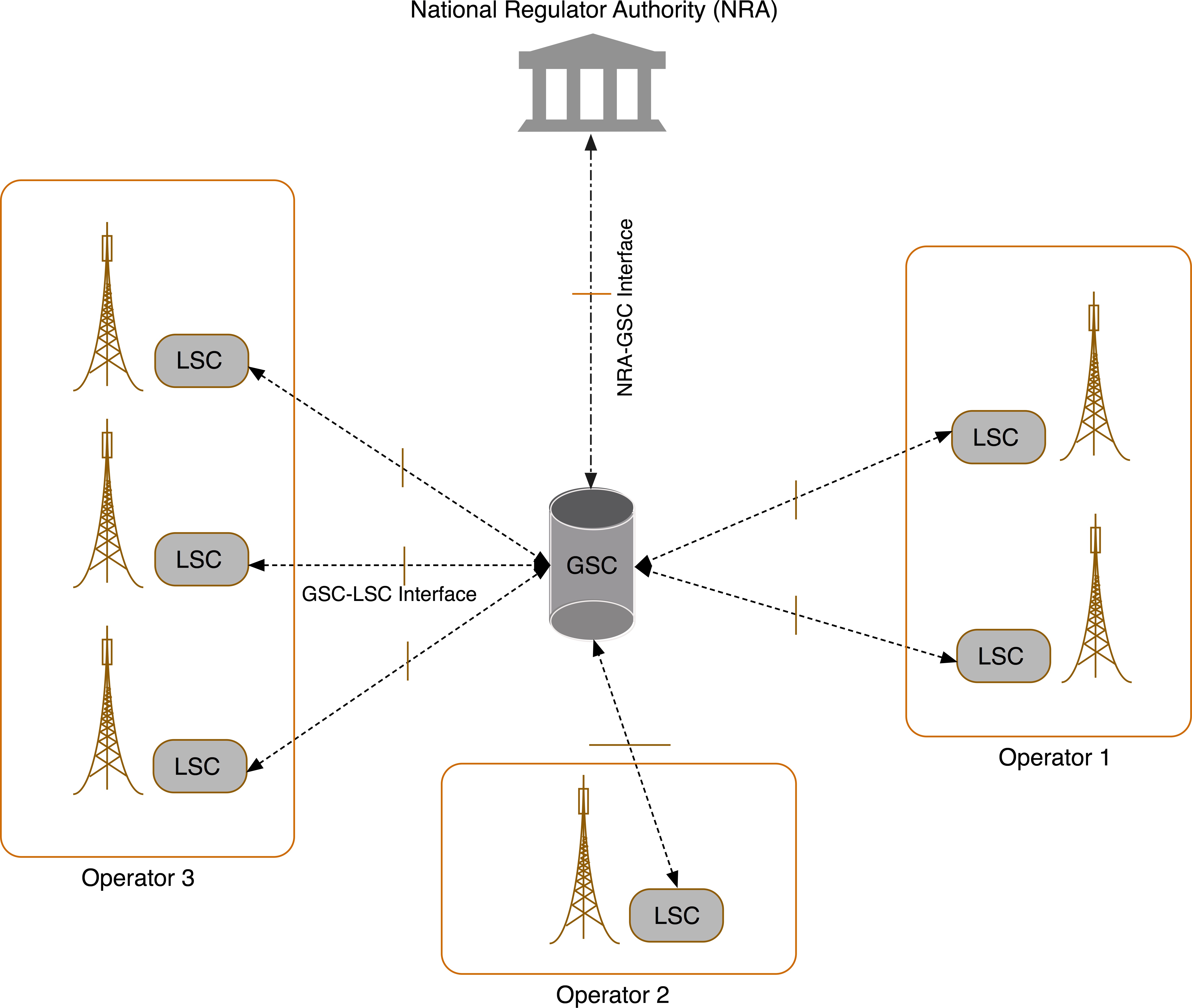}
   \caption{A typical spectrum-shared network architecture with 3 operators. One operator can have one or more base stations. An operator can be a mobile or fixed communication network. Operators with more than one operator may have one combined LSC (CLSC) where the CLSC communications with the GSC instead of individual LSC-GSC communication. }
   \label{FIG1}
\end{figure}
We consider a co-primary or horizontal spectrum sharing communications model with $N_{\rm{op}}$ operators, where  all the operators have equal right to access the spectrum. The {\it{national spectrum regulator authority (NRA)}} licences a shareable spectrum band to $N_{\rm{op}}$ operators participating in shared spectrum access, without fixed boundaries between spectrum bands of different operators. The operators coordinate their spectrum usage according certain sharing rules and mutual agreement. All the $N_{\rm{op}}$ participating operators employ orthogonal frequency division multiple access or multi-carrier waveforms \cite{Luo}, e.g., OFDMA, filter bank multi-carrier waveforms, and spectrum sharing is achieved in a coordinated way. Different operators have the flexibility to use different air-interfaces that support scalable bandwidth and flexible size of {\it{discrete Fourier transform}} (DFT). An entity called the {\it{global spectrum controller}} (GSC) carries out the coordinated spectrum sharing related tasks. The GSC can be either a virtual entity implemented in a distributed way in the base stations or an separate/independent entity in the network infrastructure. A typical shared spectrum network architecture with three operators is depicted in Fig.~\ref{FIG1}.

An operator or a shared spectrum licensee is an entity operating a {\it{mobile/fixed communication networks}} (MFCN), which holds individual rights of use to the shared spectrum resource. An operator can serve its users by one or more base stations. The GSC supports the entry and storage of shared spectrum resources availability informations, and is able to convey the related availability informations to authorized {\it{licensee shared spectrum controllers}} (LSC), and is also able to receive and store acknowledgement informations sent from the LSCs. The GSC also provides means for the NRA to monitor the operation of the shared spectrum system, and to provide the shared spectrum system with information on the {\it{Sharing Framework}} (set of sharing rules or sharing conditions for the band, information on spectrum that can be made available for shared use and the corresponding technical and operational conditions for its use) and the shared spectrum licensees. The GSC ensures that the shared spectrum system operates in conformance with the {\it{Sharing Framework}} and the licensing regime. 

LSC is located within the shared spectrum licensee's domain, and enables the shared spectrum licensee to obtain shared spectrum resource availability informations from the GSC, and to provide acknowledgment information to the GSC. The LSC interacts with the licensee's MFCN in order to support the mapping of availability informations into appropriate radio transmitter configurations, and receive the respective confirmations from the MFCN. Each base station is associated with one LSC. Multiple LSCs of the same or different shared spectrum licensee(s) are connected to one GSC.

\begin{figure}
  \centering
   \includegraphics[scale=.07]{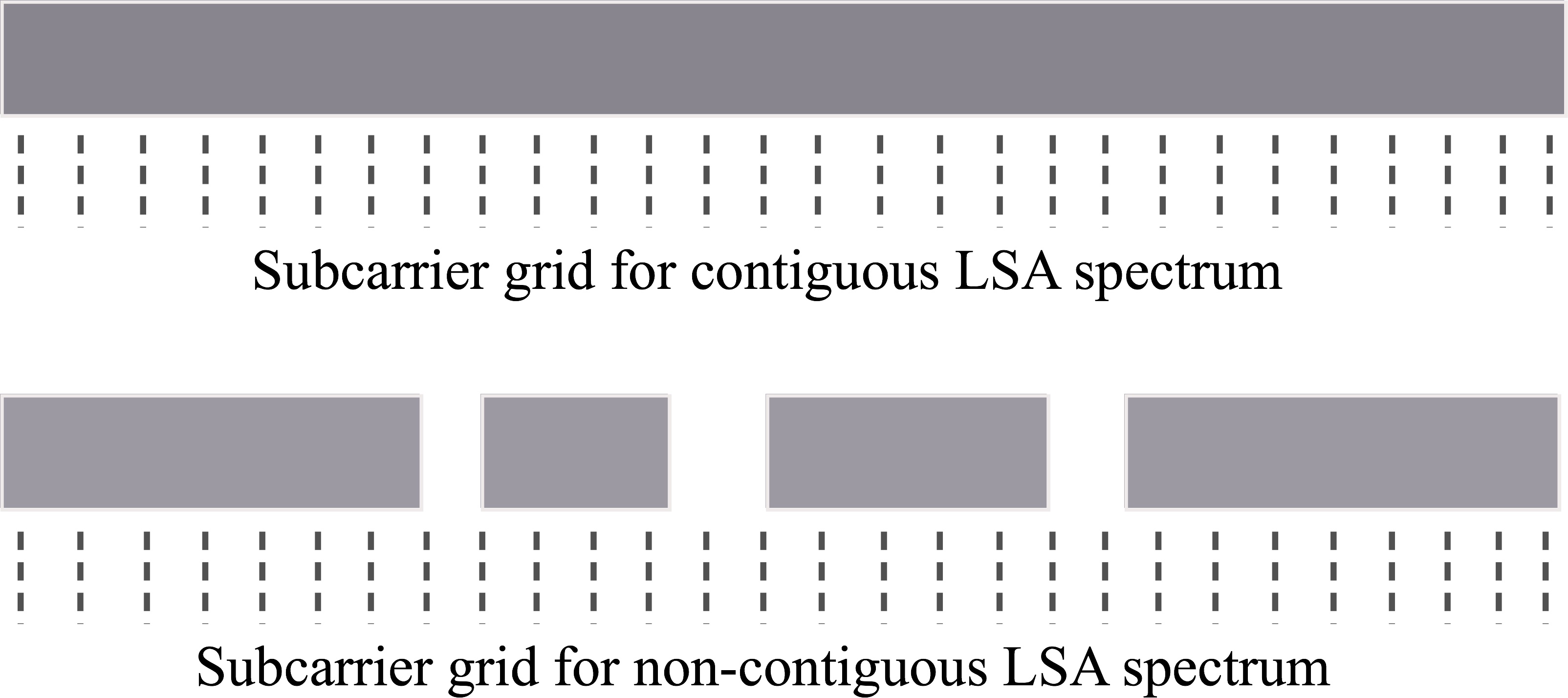}
   \caption{Common subcarrier grid, $\mathcal{S}_{\rm{grid}}$ formation for both contiguous and non-contiguous shared spectrum. }
   \label{FIG2}
\end{figure}
In our considered system model, a common subcarrier grid, $\mathcal{S}_{\rm{grid}}$ is firstly formed. The number of subcarriers $N_{\rm{sub}}$ and subcarrier spacing $\Delta_{\rm{sub}}$ depend on various parameters such as coherence bandwidth $B_{\rm{coh}}$, coherence time $T_{\rm{coh}}$, {\it{carrier frequency offset}} (CFO) tolerance $\text{CFO}_{\rm{tol}}$, Doppler frequency $f_{\rm{D}}$, the size of the smaller band of non-contiguous spectrum $B_{\rm{low}}$, if the shareable spectrum is non-contiguous, etc. The parameter $\Delta_{\rm{sub}}$, is carefully designed such that all the subcarriers in $\mathcal{S}_{\rm{grid}}$ experience flat fading. In general, $B_{\rm{coh}}$ and $T_{\rm{coh}}$ decide the maximum/minimum value of $\Delta_{\rm{sub}}$. Therefore,
\begin{equation}
\{N_{\rm{sub}},\Delta_{\rm{sub}}\}=f\left(B_{\rm{low}},B_{\rm{coh}},T_{\rm{coh}},\text{CFO}_{\rm{tol}},f_{\rm{D}}\right).
\end{equation}
A uniform common subcarrier grid is assumed, i.e., even if the shared spectrum is non-contiguous, the subcarrier grid covers the whole shared spectrum under operation. The formation of common subcarrier grid for both contiguous and non-contiguous shared spectrums is depicted in Fig.~\ref{FIG2}. 

Each operator first calculates the minimum required spectrum size in terms of Hz or the number of subcarriers, calculated based on its users' data-rate requirements and transmitting power budget. Thereupon, it sends this information along with the subcarriers' {\it{signal-to-noise ratios}} (SNRs) of its users to the GSC. The GSC associates the operators with the parameters $\left\{\rho_1, \rho_2, \cdots, \rho_{N_{\rm{op}}}\right\}$ that reflect the sharing agreement policy and quantify the operators' priority when the spectrum is shared. We consider that operators 
utilize this priority scheme in which higher priority incurs in upfront
payment to the other operators. Any operator that
desires higher priority, for instance, to guarantee more spectrum to its
customers, shall compensate the other operators by a predefined fee also
proportional to the degree of prioritization. The GSC, thereafter, determines the amount of spectrum for each operator depending on the sharing agreement policy, traffic demands and subcarrier SNRs. The common rule is that the operator who pays more, should have access to larger amount of shareable spectrum. Let $\{\Delta_{\rm{min}}^{\rm{req,1}},\cdots,\Delta_{\rm{min}}^{{\rm{req}},{N_{\rm{op}}}}\}$ denote the minimum amount of spectrums requested by the operators to satisfy their users' service requirements, calculated based on average subcarrier SNR and transmitting power constraint, where $\{\bm{q}_1,\cdots,\bm{q}_{N_{\rm{op}}}\}$ represent the sets of channel quality informations, i.e., subcarrier SNRs, calculated by different operators. The shared-spectrum allocation to different operators is obtained as

\begin{equation}
\label{main0}
 \begin{array}{*{35}{l}}
 \{{\bm{\mathcal{S}}}_1,\hspace{-.5mm}\cdots \hspace{-.5mm},{\bm{\mathcal{S}}}_{N_{\rm{op}}}\}= \\
\text{}f\left(\{\rho_1,\hspace{-.5mm}\cdots\hspace{-.5mm}, \rho_{N_{\rm{op}}}\},\{\bm{q}_1,\hspace{-.5mm}\cdots\hspace{-.5mm},\bm{q}_{N_{\rm{op}}}\},\{\Delta_{\rm{min}}^{\rm{req,1}},\hspace{-.5mm}\cdots\hspace{-.5mm},\Delta_{\rm{min}}^{{\rm{req}},{N_{\rm{op}}}}\}\right),\vspace{1.5mm} \\
\end{array}
\end{equation}
where $\{{\bm{\mathcal{S}}}_1,\cdots, {\bm{\mathcal{S}}}_{N_{\rm{op}}}\}$  are the sets of subcarriers assigned to spectrum sharing operators and ${\bm{\mathcal{S}}}_1\cup{\bm{\mathcal{S}}}_2\cup\cdots\cup{\bm{\mathcal{S}}}_{N_{\rm{op}}}=\mathcal{S}_{\rm{grid}}$. Note that the subcarriers in ${\bm{\mathcal{S}}}_n$ can be from a small fragment (a contiguous band) of shared spectrum or can be scattered over the whole shared spectrum.

Let us consider that spectrum sharing operator $n$ supports $K_n$ non-cooperative users with a single receiving antenna each. The data transmissions of different users are assumed to be subject to slowly-varying, independent frequency-selective Rayleigh fading. Perfect channel state information is assumed to be available and a non-sharing subcarrier allocation scheme is considered, i.e., a subcarrier can be allocated to a single user only. The data transmission is subject to regulated maximum transmitting power constraint, $P^{(n)}_{\rm{Max}}$. Let us consider that $||\mathcal{S}_n||=L_n$ and $B_n=\Delta_{\rm{sub}}L_n$ is the total frequency bandwidth allocated to operator $n$. Then the capacity achieved by user $k$ of operator $n$ when transmitting data over subcarrier $l$ is given by
\begin{equation}
\label{rate}
r^{(n)}_{k,l}={\rm{log_2}}\left( 1+p^{(n)}_{k,l}h^{(n)}_{k,l} \right),
\end{equation}
with $h^{(n)}_{k,l}= |z^{(n)}_{k,l}|^2 /{\sigma}^2_{n}$, where $z^{(n)}_{k,l}$ defines the frequency gain on subcarrier $l$ of user $k$ and $||\bm{x}||$ denotes the cardinality of $\bm{x}$. ${\sigma}^2_{n}=N_0 B_n/L_n$ is the variance of additive white Gaussian noise (AWGN) over subcarrier $l$, where $N_0$ is the noise power spectral density. The quantity $h^{(n)}_{k,l}=|z^{(n)}_{k,l}|^2/\left(N_0 \frac{B_n}{L_n}\right)$ is defined as the effective SNR on subcarrier $l$ allocated to user $k$ of operator $n$. $p^{(n)}_{k,l}$ is the amount of power allocated to user $k$ corresponding to subcarrier $l$. The total rate achieved by user $k$ of operator $n$ is given by $R^{(n)}_k=\sum\limits_{l=1}^{L_n}{{{{c}}^{(n)}_{k,l}}{{r}^{(n)}_{k,l}}}$, where ${{{c}}^{(n)}_{k,l}}$ is the subcarrier assignment function. ${{{c}}^{(n)}_{k,l}}=1$ refers to the subcarrier allocation in which user $k$ of operator $n$ is assigned with subcarrier $l$. If subcarrier $l$ is not assigned to user $k$, ${{{c}}^{(n)}_{k,l}}$ is equal to 0.

\section{Solutions for Inter-operator Spectrum Sharing}

In our proposed inter-operator spectrum sharing approaches, spectrum sharing among the operators is transformed into an optimization program which maintains a fair priority requirements based on the sharing policy and traffic demands of the operators while maximizing the total system throughput. The spectrum allocation among the operators is performed in such a way that coarsely fulfills the relationship as follows 
\begin{equation}
\label{mainx10}
\frac{||\mathcal{S}_i||}{||\mathcal{S}_j||}\approx\frac{\rho_i^{\rm{act}}}{\rho_j^{\rm{act}}}\hspace{3mm}\forall i,j\in \mathcal{N},i\ne j,
\end{equation}
where $\mathcal{N}\triangleq\left\{1,2,\cdots,N_{\rm{op}}\right\}$ and $\sum_{n=1}^{N_{\rm{op}}} \rho_n^{\rm{act}}=1$. The relationship defined in \eqref{mainx10} states that the amount of spectrum resources allocated to the operators are proportional to each other, and $\rho_n^{\rm{act}}$ defines the active priority measure of operator $n$. Note that $\rho_i, \forall i, i\in \mathcal{N}$ are the original priority measures of the operators depending solely on sharing rules and mutual agreement. While $\rho_i^{\rm{act}}, \forall i, i\in \mathcal{N}$ are the active priority measures calculated considering additionally the current traffic demands, and $\rho_i^{\rm{act}}, \forall i, i\in \mathcal{N}$ decide the final spectrum allocation. The values of $\rho_i^{\rm{act}}, \forall i, i\in \mathcal{N}$ may or may not be equal to the values of $\rho_i, \forall i, i\in \mathcal{N}$.

\subsection{Subcarrier gain based spectrum sharing} 
\begin{figure*}
  \centering
\includegraphics[scale=.05]{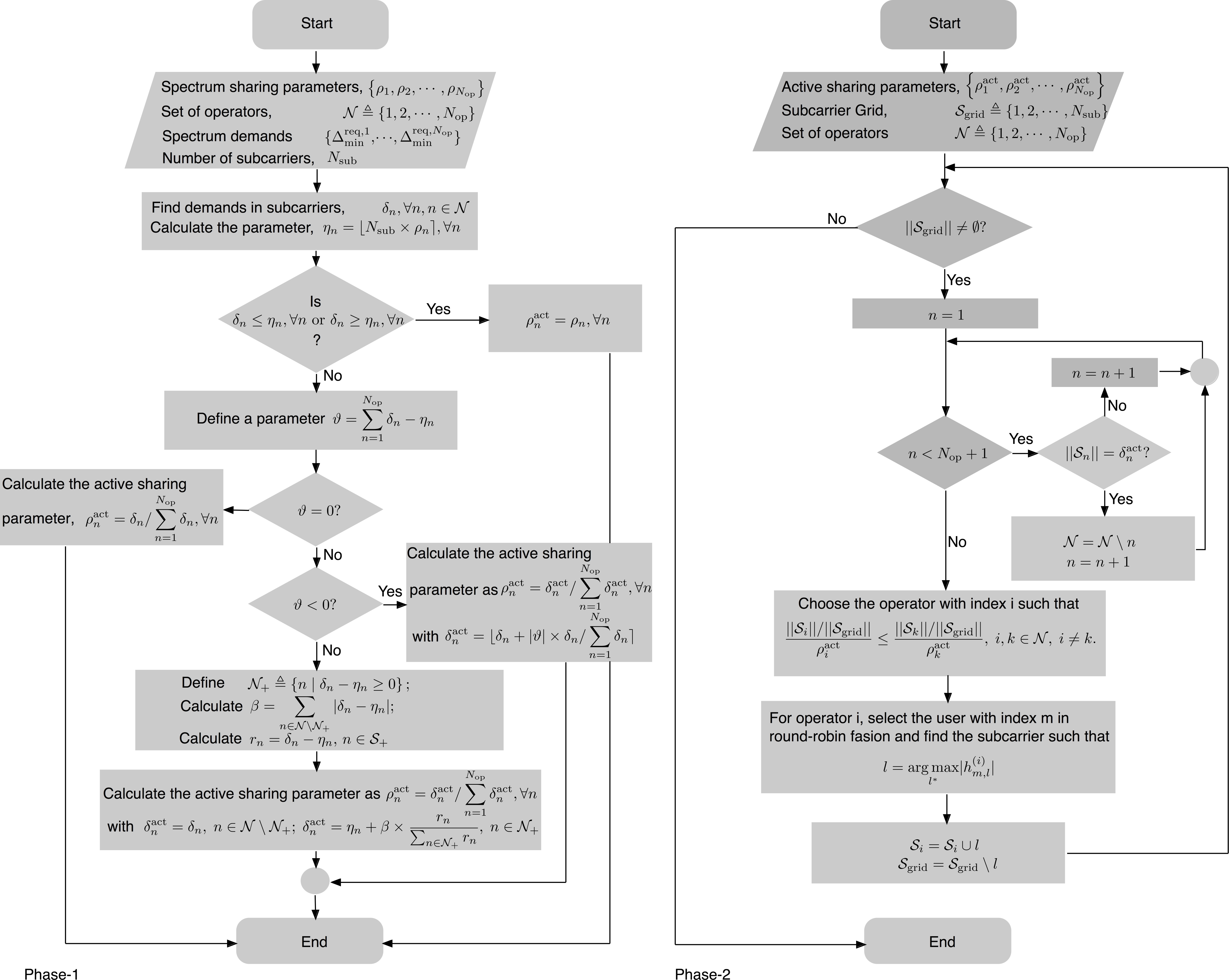}
   \caption{Proposed subcarrier gain based resource sharing in shared spectrum access communication systems. Phase-1 corresponds to the process for calculating the active set of sharing parameters $\left\{\rho_1^{\rm{act}},\rho_2^{\rm{act}},\cdots,\rho_{N_{\rm{op}}}^{\rm{act}}\right\}$, and Phase-2 corresponds to the resource allocation process among the spectrum sharing operators based on calculated $\left\{\rho_1^{\rm{act}},\rho_2^{\rm{act}},\cdots,\rho_{N_{\rm{op}}}^{\rm{act}}\right\}$ in Phase-1. Note that for fragmentation based spectrum sharing, Phase-1 remains the same. }
   \label{FIG12}
\end{figure*}

The proposed subcarrier based inter-operator spectrum sharing approach is iterative in nature, and opts to fairly and flexibly allocate the available shareable spectrum among the operators. The subcarriers allocated to any operator scatter over whole subcarrier grid.
We can express the spectrum allocation problem as
\begin{equation}
\label{main0}
 \begin{array}{*{35}{l}}
\underset{\{\mathcal{S}_n\}}{\mathop{\max }}\,\sum\limits_{n=1}^{{N_{\rm{op}}}}\sum\limits_{k=1}^{K_n}R_k^{(n)}\vspace{1.5mm} \\
\text{}\text{subject to} \hspace{1mm}\frac{||\mathcal{S}_i||}{||\mathcal{S}_j||}\approx\frac{\rho_i^{\rm{act}}}{\rho_j^{\rm{act}}}\hspace{3mm}\forall i,j\in \mathcal{N},i\ne j.\vspace{1.5mm} \\
\text{}\hspace{15mm}\mathcal{S}_1\cup \mathcal{S}_2\cup\cdots\cup\mathcal{S}_{N_{\rm{op}}}=\mathcal{S}_{\rm{grid}}. \vspace{1.5mm} \\
\end{array}
\end{equation}
The GSC calculates an active set of $\left\{\rho_1, \rho_2, \cdots, \rho_{N_{\rm{op}}}\right\}$, defined as $\left\{\rho_1^{\rm{act}},\rho_2^{\rm{act}},\cdots,\rho_{N_{\rm{op}}}^{\rm{act}}\right\}$ before entering into the actual iterative process. The determination of this active set and the iterative spectrum allocation process is provided in Fig.~\ref{FIG12}. The GSC calculates the active set of sharing parameters by following the process described in {\bf{Phase-1}} and finally the spectrum allocation among the operators is carried out by following the process illustrated in {\bf{Phase-2}}.

The spectrum allocation problem with the operators' desired amount of spectrum and preferred sets of subcarriers is transformed into an optimization program which maintains a proportional spectrum fairness requirements among the operators. The amount of spectrum and subcarriers are assigned to the operators in such a way that coarsely/closely fulfills the relationship given in \eqref{main0}. The advantage of introducing these normalized priority spectrum measures $\left\{\rho_1^{\rm{act}},\rho_2^{\rm{act}},\cdots,\rho_{N_{\rm{op}}}^{\rm{act}}\right\}$ is that we can explicitly control the spectrum allocation ratio among the operators during an iterative allocation process. At any particular iteration, the operator $i$ has the opportunity to get assigned with a subcarrier that has the highest SNR over all of its users if it complies with the condition: 
\begin{equation}
\frac{||\mathcal{S}_i||/||\mathcal{S}_{\rm{grid}}||}{\rho_i^{\rm{act}}}\le \frac{||\mathcal{S}_k||/||\mathcal{S}_{\rm{grid}}||}{\rho_k^{\rm{act}}},\hspace{1mm} i,k\in \mathcal{N},\hspace{1mm}i\neq k. 
\end{equation}
Hence, this proposed spectrum allocation scheme assigns subcarriers to the operators depending on their desired amount of spectrum and the sharing policy.

A set of predefined rules is followed by the GSC for dynamic and fair allocation of shared-spectrum. The amount of shared-spectrum allocated to any operator $n$, $\delta_n$ (calculated based on $P^{(n)}_{\rm{Max}}$ and average subcarrier SNR $\bar{h}^{(n)}$) scales with its traffic demand under some fairness measures. If the desired amounts of shared-spectrum in terms of number of subcarriers, $\delta_n, \hspace{1mm}\forall n, \hspace{1mm}n\in \mathcal{N}$ are $\ge$ or $\le$ their actual priority amounts $\eta_n, \hspace{1mm}\forall n, \hspace{1mm}n\in \mathcal{N}$ corresponding to the sharing agreement ($\left\{\rho_1, \rho_2, \cdots, \rho_{N_{\rm{op}}}\right\}$), the GSC just allocates the subcarriers proportionally according to the actual sharing agreement policy, i.e., $\rho_n^{\rm{act}}=\rho_n, \forall n$.

When for some operators, $\delta_i<\eta_i, i\in\mathcal{N}$, and for the rest of the operators $\delta_j\ge\eta_j, j\in\mathcal{N}$ with $i\neq j$, then if $\sum_{n=1}^{{N_{\rm{op}}}}\delta_n-\eta_n\le0$, the GSC allocates the spectrum according to the desired amount of spectrum irrespective of the sharing agreement as it will not violate the agreement. However, if $\sum_{n=1}^{{N_{\rm{op}}}}\delta_n-\eta_n>0$, the GSC allocates the additional spectrum due to the operators who have $\delta_n<\eta_n$ among the operators with $\delta_n>\eta_n$ proportionally depending on the values of $\delta_n-\eta_n$. Furthermore, for each operator, the subcarrier assignment is performed in such a way that maintains fairness among its users, i.e., each user gets equal opportunity in a round-robin manner to get assigned with its best subcarriers. Finally, after the GSC finds the subcarrier sets $\left\{\mathcal{S}_1,\mathcal{S}_2,\cdots, \mathcal{S}_{N_{\rm{op}}}\right\}$, it notifies all the spectrum sharing operators. Since all the network operators have access to the whole spectrum band and the users' device has the capability to tune to the whole band, each operator can give access to any user subscribed to one of the sharing network operators. However, in order to enable such spectrum sharing, infrastructure sharing and high coordination is required among the operators. Software defined networking and network function virtualization may be the viable solutions for such sharing in order to enable the system obtaining the benefit from multiuser diversity across the operators' domain.

Note that this subcarrier based spectrum allocation process is iterative and thus requires significantly more time to find the allocation. If the operating environment is such that the channel is highly frequency-selective, then ${N_{\rm{sub}}}$ tends to become larger since the subcarrier spacing will become shorter, which in turn, will increase the computation time. Note that in the current study we have considered that the finest resource granularity for transmission is one subcarrier. However, the finest resource
granularity can also be one resource block as in the case of long term evolution-advanced (LTE-A), which contains a group of successive subcarriers.

\subsection{Fragmentation based spectrum sharing}
In order to reduce the computation time, the GSC can perform fragmentation based shared spectrum allocation. 
For fragmentation based spectrum sharing we follow the same procedures for obtaining the active priority measures, i.e., {\bf{Phase-1}} remains the same.

In fragmentation based spectrum sharing, the operators have the option to inform the GSC about the favorable fragments they like to operate on by sending an extra variable $\alpha$. The range of values in $\alpha$ depends on the number of operators participating in the spectrum sharing process. For example, if there are only two operators, $\alpha$ can be binary. If any operator wants to transmit on the lower end of the spectrum, it sends $0$, or $1$, otherwise. The same goes for the second operator. If there are 3 operator, $\alpha$ is of 2 bits size while $00$ points to the lower end and $11$ points to the upper end of the spectrum. Any operator favors one fragment over other fragments in the shared frequency band depending on the types of applications, e.g., short-range communications, the technologies it operates on and channel propagation characteristics, and its achievable capacity on the fragment. The GSC also takes these features under consideration when it allocates the fragments to different operators.

When two or more operators request for the same fragment, the GSC prioritizes one over the other by judicious evaluation of the above mentioned features. If all the operators offer the same type of applications and have equal priority, then the contention is solved by random selection of one of the operators. When similar contention occurs again in future, the GSC performs the same random selection by ignoring the operator that was selected in the previous contention period.
When fragmentation based spectrum sharing is employed, each operator can have independent network deployment. Each operator can independently adjust its own transmission frame structure in accordance with the use case, traffic type etc. The operators have also the flexibility to change the number of subcarriers thus the subcarrier spacing, $\Delta_{\rm{sub}}$ within their fragments.

As an example let us consider that there are only two operators participating in the shared spectrum access communications, and the available spectrum for sharing in contiguous. If both the operators have same applications types, e.g., cellular communications and $\rho_1=\rho_2=0.5$ with $\alpha_1=0,\hspace{1mm}\alpha_2=1$, respectively. The GSC calculates the active priority measures $\rho_1^{\rm{act}}$ and $\rho_1^{\rm{act}}$ in accordance with the traffic demands from the operators and sharing rules. The GSC then partitions the shareable spectrum into two fragments with respect to  $\rho_1^{\rm{act}}$ and $\rho_1^{\rm{act}}$, and allocates the lower end fragment of size $\delta_1^{\rm{act}}\Delta_{\rm{sub}}$ Hz to operator 1 and the upper end fragment of size $\delta_2^{\rm{act}}\Delta_{\rm{sub}}$ Hz to operator 2. However, if $\alpha_1$ and $\alpha_2$ are equal, then the GSC randomly assigns the ends of the spectrum to the operators. If the communications scenario is such that the type of applications and technologies one of the operators operates on, e.g., shorter-range communications, prefers higher end fragment to upper end fragment, the GSC will probably allocate the higher end fragment to that particular operator, even if the other operator seeks the same end. Therefore, fragmentation based spectrum sharing is a situation-specific mechanism. 

As we have already mentioned in previous sections, the GSC can allocate the shared-spectrum among the operators based on fragmentation depending on the operators' desired minimum fragment size and channel quality information. A fragment is defined as a contiguous band and its bandwidth must be an integer multiple of the subcarrier bandwidth $\Delta_{\rm{sub}}$  and complies with $B_{\rm{low}}$ and other system parameters. It is worth mentioning that if the fragmentation based spectrum sharing is employed, then each operator has the flexibility to employ independent radio interface, flexible DFT size, etc. The work in \cite{Mirza5} considers a spectrum spectrum sharing scenario where different operators employ different radio interfaces, i.e, waveforms. It is also possible to allocate different fragments from non-contiguous bands if the radio interface supports it \cite{McMenamy}. If the operators are not synchronized in a way that operators of adjacent spectrum do not transmit at the same time, guard bands are created around each fragment  to protect other operators from its out-of-band emission. The minimum size of the fragments depends on the desired guard band overhead, which in turn, depends on the pulse shape being used for data modulation.

Each operator can claim a minimum fragment size to limit the overall guardband overhead given as $\frac{B_{\rm{guard}}}{B_{\rm{frag}}}$, where $B_{\rm{guard}}$ is total amount of spectrum belonging to all the guardbands and $B_{\rm{frag}}$ is the total amount of frequency spectrum belonging to all usable fragments. If the fragment size is too small, the guardband overhead would be too large. This is calculated based on its waveform especially the out-of-band emission level. From that the operator calculates the required guardband and considering the guardband size, the operator then claims a minimum fragment size. The GSC performs the fragmentation according to the traffic loads of each operator and may also depend on the preferred fragments of each operator.

Each operator transmits a signal in the allocated spectrum fragments by activating and deactivating subcarriers of the signal, i.e., only the subcarriers within the allocated spectrum fragments are activated, while others are not. Since different operators do not necessarily have accurate mutual synchronization, the out-of-fragment radiation power of the signals of each operator has to be taken into account, which causes interference to the other operators. We also consider accurate inter-operator synchronization that can be realized in downlink if the spectrum sharing operators share the radio access network or through GPS modules used in the base stations. In this proposed system model, each operator can transmit data to its own serving users independently without creating interference (assuming perfect synchronization among the operators or adequate guardbands between the fragments if fragmentation based spectrum sharing is employed) to the users served by other operators. 

Both of the proposed inter-operator spectrum sharing solutions are suboptimal. In subcarrier or resource block based resource sharing optimization, the solution aims at optimizing the total system throughput and resource allocation is obtained through an iterative process instead of optimal exhaustive search in order to reduce the computational burden, and ensures fairness among the operators as well as the users. In fragmentation based spectrum sharing optimization, GSC judiciously evaluates the sharing rules and the informations received from the operators, and allocates the favorable spectrum fragments accordingly. Note that GSC does not follow any strict mathematical process, instead it takes dynamic and situation-specific measures to decide the fragment allocation.

\section{Solution for Intra-operator Resource Allocation}
Soon after receiving the information about spectrum allocations from the GSC, each operator performs intra-operator resource allocation for the users with dissimilar service requirements. In practice, the regulatory scenario enforces a total transmitting or radiated power constraint. Therefore, the base stations of each operator work under maximum transmitting power constraint while satisfying its own users' service requirements. Note that there can be several base stations under one operator. In this study, we consider that each operator has only one base station to make the analysis simple and straightforward. Note that it is enough to consider  resource allocation optimization for any particular operator since base station can transmit data to its own serving users independently without creating interference (with perfect synchronization or adequate guardbands between the fragments) to the users served by other operators.

  The main objective of resource allocation optimization in this section is to perform efficient resource allocation for users with DC and NDC service requirements under total transmitting power constraint. Let us consider the resource allocation for operator $n$, for which $||\mathcal{S}_n||=L_n$, and $K_1$ users out of $K_n$ users belong to NDC service requirements while $K_n-K_1$ users require DC services under the maximum transmitting power constraint $P^{(n)}_{\rm{Max}}$. For notational brevity, we ignore the operator index $n$ in the following presentation of the paper. Let us define $\mathcal{K}$ and $\mathcal{L}$ as the sets of $\left\{1,2,\cdots,K\right\}$ and $\left\{1,2,\cdots,L\right\}$, respectively. Now, the resource allocation optimization problem can be cast as 0-1 optimization problem, which is given by 
\begin{equation}
\label{main2}
 \begin{array}{*{35}{l}}
\underset{\{{{c}_{k,l}},{{p}_{k,l}}\}}{\mathop{\max }}\,\sum\limits_{k=1}^{K_1}{\sum\limits_{l=1}^{L}{{{c}_{k,l}}{r}_{k,l}}}\vspace{1.5mm} \\
\vspace{2mm}
\text{}\text{subject to}  \vspace{1.5mm} \\
\text{}\hspace{1mm}\text{C1: }{{c}_{k,l}}\in \left\{ 0,1 \right\},\hspace{8mm}\forall k,l,\hspace{1mm}k\in\mathcal{K},\hspace{1mm}l\in\mathcal{L} \vspace{1.5mm} \\
\text{}\hspace{1mm} \text{C2: }\sum\limits_{k=1}^{{K}}c_{k,l}=1,\hspace{9mm} \forall l,,\hspace{1mm}l\in\mathcal{L}\vspace{2mm} \\
\text{}\hspace{1mm} \text{C3: }{{p}_{k,l}}\ge 0,\hspace{13.5mm}\forall k,l,\hspace{1mm}k\in\mathcal{K},\hspace{1mm}l\in\mathcal{L} \vspace{2mm} \\
\text{}\hspace{1mm} \text{C4: }\sum\limits_{k=1}^{{K}}{\sum\limits_{l=1}^{L}{{{c}_{k,l}}{{p}_{k,l}}\le {P_{\rm{Max}}}}} \vspace{1.5mm} \\
\text{}\hspace{1mm} \text{C5: }\sum\limits_{l=1}^{L}{c}_{k,l}{r}_{k,l} \ge R_k^{\rm{target}}, \hspace{1mm}k=K_1+1,\cdots, {K} 
\end{array}
\end{equation}
where, $R_k^{\rm{target}}$ is the desired throughput of DC user $k$. A solution to this problem looks for a partition of groundset $\{1,2,\cdots,L\}$ into ${K}$ subsets such that the measure associated with the subsets, $P_{\rm{Max}}$ fulfills some bounds. Note that the problem in \eqref{main2} is nonlinear since we have multiplication of variables in constraints and in the objective function. Also, we need to deal with the integer variables because of non-divisibility of resources (subcarriers), thus it is an nonconvex MINLP. Nonlinear constraints are more difficult to handle. Therefore, it is very advantageous to incorporate linear constraints. In our proposed solution, we restrict our optimization model to contain only linear constraints.  In the following, we formulate several linearization approaches to transform \eqref{main2} into a convex program, which is easier to solve than solving an MINLP since combining both nonlinearity and integrality can lead the MINLP to be undecidable \cite{Jeroslow}.

\begin{figure}
  \centering
  \includegraphics[scale=.08]{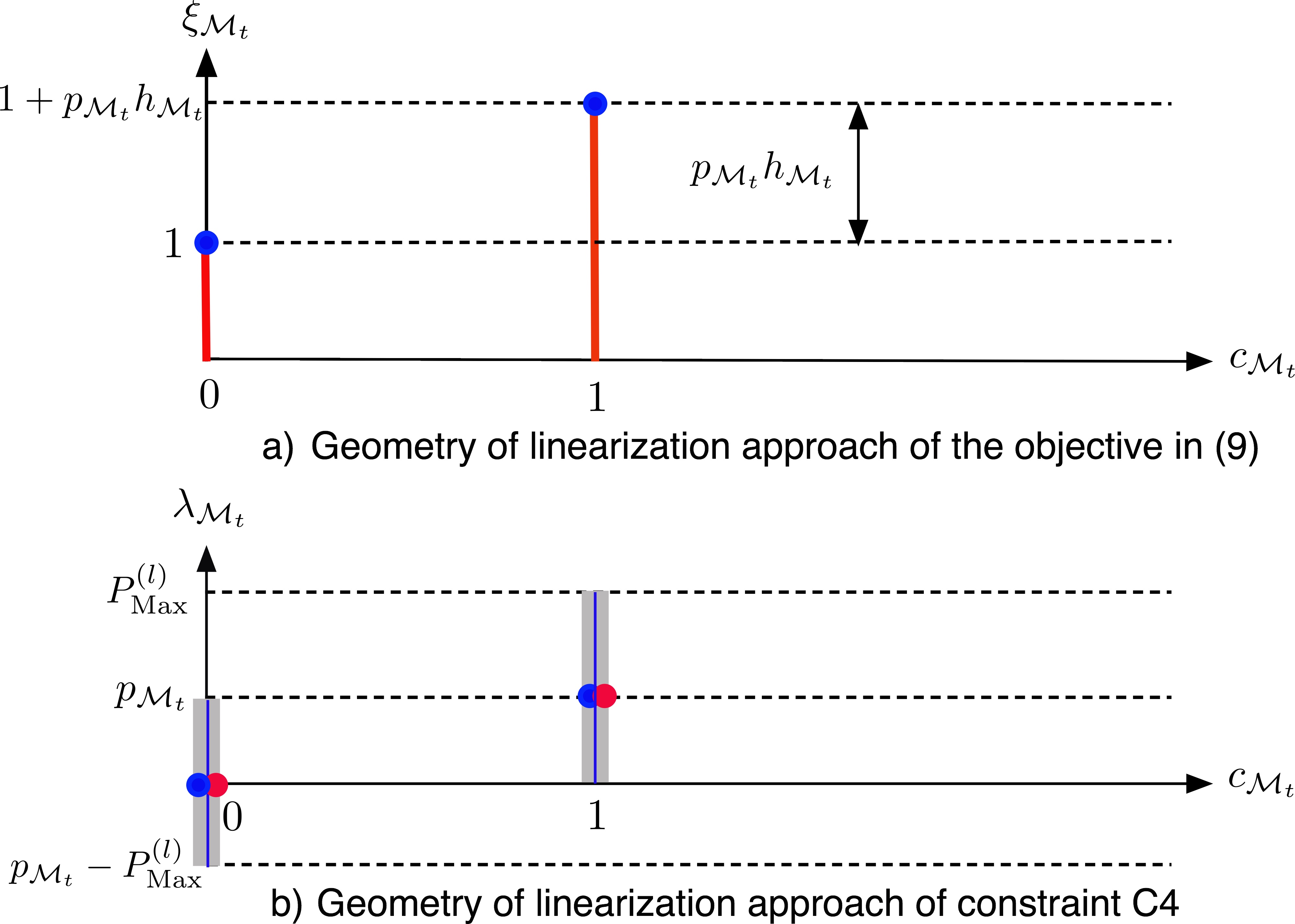}
   \caption{Geometry for linearization approach of the objective function and constraint C4 in \eqref{main3}. }
   \label{FIG3}
   \end{figure}

For notational brevity and simplified analysis, we define a set $\mathcal{M}$ that contains all the tuples of the indices, i.e., $\{k,l\}$ as
$\Big\{\{1,1\},\cdots\hspace{-.5mm}, \{1,L\},\cdots\hspace{-.5mm}, \{k,1\},\cdots\hspace{-.5mm}, \{k,L\},\cdots\hspace{-.5mm},\{{K},\hspace{-.3mm}1\}$ $,\cdots\hspace{-.5mm}, \{{K},L\}\Big\},$
where $\mathcal{M}_t$ denotes the $t$th tuple in $\mathcal{M}$. The corresponding indices in the $t$th subset are defined as  $\{k,l\}=\Big\{\lceil{t/L}\rceil, t-\big((\lceil{t/L}\rceil-1)L\big)\Big\}$, where $\lceil \cdot \rceil$ is the ceiling operator. Similarly, we vectorize the elements $h_{k,l}$, $c_{k,l}$ and $p_{k,l}, \hspace{1mm}\forall k,l$, in $\bm{h}$, $\bm{c}$ and $\bm{p}$, respectively, in accordance with the sequences of the indices in $\mathcal{M}$. 
Now, we can rewrite the objective function as
 \begin{equation}
\label{notmain1}
 \begin{array}{*{35}{l}}
 \underset{\left\{{{c}_{\mathcal{M}_t},{p}_{\mathcal{M}_t}}\right\}}{\mathop{\mathrm{max} }}\,\hspace{-1mm}\sum\nolimits_{t=1}^{K_1L}{{{c }_{\mathcal{M}_t}}{{\log }_{2}}(1+{{p }_{\mathcal{M}_t}}{{h}_{\mathcal{M}_t}})}={{\log }_{2}} \left(\underset{\left\{{{c}_{\mathcal{M}_t},{p}_{\mathcal{M}_t}}\right\}}{\mathop{\mathrm{max} }}\,\hspace{-1mm}\prod\nolimits_{t=1}^{K_1L}\cdots\right.\\
\hspace{60mm} {{{(1+{{p }_{\mathcal{M}_t}}{{h}_{\mathcal{M}_t}})}^{{{c}_{\mathcal{M}_t}}}}}\Bigg)\\
 \end{array}
\end{equation}
Now,  due to the monotonicity of logarithmic function, i.e., if $x>y,\hspace{1mm} {\rm{log}}_2(x)>{\rm{log}}_2(y)$, the log function in the subsequent optimization problems can be eliminated. We can rewrite the objective function in \eqref{notmain1} as $\underset{\{{{c}_{\mathcal{M}_t}},{{p}_{\mathcal{M}_t}}\}}{\mathop{\max }}\,\hspace{3mm} \prod\limits_{t=1}^{K_1L}{(1+{p}_{\mathcal{M}_t}{h}_{\mathcal{M}_t})}^{{c}_{\mathcal{M}_t}}$, and consequently, the optimization problem in \eqref{main2} can be expressed as
  \begin{equation}
\label{main3}
 \begin{array}{*{35}{l}}
\underset{\{{{c}_{\mathcal{M}_t}},{{p}_{\mathcal{M}_t}}\}}{\mathop{\max }}\,\hspace{1mm} \prod\limits_{t=1}^{K_1{L}}{(1+{p}_{\mathcal{M}_t}{h}_{\mathcal{M}_t})}^{{c}_{\mathcal{M}_t}}\vspace{2mm} \\
\text{}\text{subject to} \vspace{2mm} \\
\text{}\hspace{0mm}\text{C1:} \hspace{1mm}{c}_{\mathcal{M}_t}\in\{0,1\}\hspace{25mm}\forall t,\hspace{1mm}t=1,\cdots,{K}{L} \vspace{2mm} \\
\text{}\hspace{0mm}\text{C2:}\hspace{1mm}\sum_{\begin{matrix} t\in\{l,l+k{L}\}\\k=1,\cdots, {K}\end{matrix}}{c}_{\mathcal{M}_t}=1, \hspace{3.8mm}\forall l,\hspace{1mm} l=1,\cdots,{L} \vspace{2mm} \\
\text{}\hspace{0mm}\text{C3:}\hspace{1mm} {p}_{\mathcal{M}_t}\ge 0, \hspace{25mm}\hspace{4mm}\forall t,\hspace{2mm}t=1,\cdots, {K}{L} \vspace{2mm} \\
\text{}\hspace{0mm}\text{C4:}\hspace{1mm} \sum\limits_{t=1}^{{K}{L}}{c}_{\mathcal{M}_t}{p}_{\mathcal{M}_t}\le P_{\rm{Max}} \vspace{2mm} \\
\text{}\hspace{0mm}\text{C5:}\hspace{1mm} \sum\limits_{t=K_1{L}+1+(k-1){L}}^{K_1{L}+{L}+(k-1){L}}\hspace{-8mm}{c}_{\mathcal{M}_t}{\rm{log_2}}(1+\hspace{-.5mm}{p}_{\mathcal{M}_t}{h}_{\mathcal{M}_t})\hspace{-.5mm}\ge\hspace{-.5mm} R_k^{\rm{target}}{L}, \hspace{1mm} k=K_1+1,\cdots\hspace{-.5mm}, {K}  
\end{array}
\end{equation}
The objective function is still nonlinear in its current form. However, note that since ${c}_{\mathcal{M}_t}\in\{0,1\}$, we have the flexibility to transform each multiplicative term in the objective function, $(1+{p}_{\mathcal{M}_t}{h}_{\mathcal{M}_t})^{{c}_{\mathcal{M}_t}}$ as
\begin{equation}
\label{notmain2}
(1+{p}_{\mathcal{M}_t}{h}_{\mathcal{M}_t})^{{c}_{\mathcal{M}_t}}=1+{c}_{\mathcal{M}_t}{p}_{\mathcal{M}_t}{h}_{\mathcal{M}_t}.
\end{equation}
Obviously, this isn't true for all ${c}_{\mathcal{M}_t}$, it is true only for binary ${c}_{\mathcal{M}_t}$, which is the case here. We take advantage of this transformational relationship in the linearization process. For each $t$, we create a new variable ${\xi}_{\mathcal{M}_t}$, and then add the constraints as given below
\begin{equation}
{\xi}_{\mathcal{M}_t}\le 1+{\rm{min}}\{{p}_{\mathcal{M}_t},P_{\rm{Max}}{c}_{\mathcal{M}_t}\}{h}_{\mathcal{M}_t}.
\label{equation1}
\end{equation}
Consider the first case ${c}_{\mathcal{M}_t}=0$, which means the value of $(1+{p}_{\mathcal{M}_t}{h}_{\mathcal{M}_t})^{{c}_{\mathcal{M}_t}}$ should be 1. The linear constraint ${\xi}_{\mathcal{M}_t}\le 1+{\rm{min}}\{{p}_{\mathcal{M}_t},P_{\rm{Max}}{c}_{\mathcal{M}_t}\}{h}_{\mathcal{M}_t}$ in \eqref{equation1} forces ${\xi}_{\mathcal{M}_t}$ to be $\le$ 1. Now, let us consider the case ${c}_{\mathcal{M}_t}=1$, now the quantity should be equal to $1+{p}_{\mathcal{M}_t}{h}_{\mathcal{M}_t}$. Similarly, the constraint in  \eqref{equation1} enforces the quantity ${\xi} _{\mathcal{M}_t}$ to be $\le$ $1+{p}_{\mathcal{M}_t}{h}_{\mathcal{M}_t}$. The reason for putting a `$\le$' sign in \eqref{equation1} instead of the `$=$' sign is to obtain a convex feasible region since a constraint in the form of \textit{ ``linear/affine $\ge$ concave"} generates a non-convex feasible region, which is a major problem. However, at the optimum, the inequality turns out to be an equality. The geometry of the linearization process of constraint C3 is depicted in Fig.~\ref{FIG3}a. The blue circles represent the exact or expected values of the variable ${\xi}_{\mathcal{M}_t}$ for two different cases (${c}_{\mathcal{M}_t}=0$ and ${c}_{\mathcal{M}_t}=1$). The red lines show the range of values the variable can take when linearization is employed.
   
Furthermore, maximization of the new objective function, $\prod\limits_{t=1}^{K_1{L}}{{\xi}_{\mathcal{M}_t}}$ and maximization of $ \left(\prod\limits_{t=1}^{K_1{L}}{{\xi}_{\mathcal{M}_t}}\right)^{1/K_1{L}}$ will give the same values for the optimization variables. So, we can recast the optimization problem in \eqref{main3} as the following
\begin{equation}
\label{main4}
 \begin{array}{*{35}{l}}
\underset{\{{{c}_{\mathcal{M}_t}},{\xi}_{\mathcal{M}_t}{{p}_{t}}\}}{\mathop{\max }}\,\hspace{2mm} \left(\prod\limits_{t=1}^{K_1{L}}{{\xi}_{\mathcal{M}_t}}\right)^{1/K_1{L}}\vspace{3mm} \\
\text{}\text{subject to} \vspace{3mm} \\
\text{}\hspace{5mm}\text{C6:}\hspace{6mm}{\xi}_{\mathcal{M}_t}\le 1+{\rm{min}}\{{p}_{\mathcal{M}_t},P_{\rm{Max}}{c}_{\mathcal{M}_t}\}{h}_{\mathcal{M}_t} \vspace{2mm} \\
\end{array}
\end{equation}
along with the constraints C1-C5 in \eqref{main3}. Now, the objective of the optimization problem in \eqref{main4} becomes a concave function.

Now, let us turn our focus to the linearization of the constraint C4 in \eqref{main3}. Here, ${c}_{\mathcal{M}_t}$ (binary integer) and ${p}_{\mathcal{M}_t}$ (continuous) are variables in this mathematical program, and we have to deal with their product ${c}_{\mathcal{M}_t}{p}_{\mathcal{M}_t}$. If both the variables were continuous, we would have ended up having a quadratic problem, which may create issues with convexity if the quadratic terms appear in constraints. However, constraint C4 is special in the sense that at least one of ${c}_{\mathcal{M}_t}$ and ${p}_{\mathcal{M}_t}$ is binary, and the other variable is bounded. Under these assumptions, the product ${c}_{\mathcal{M}_t}{p}_{\mathcal{M}_t}$ can be linearized by introducing new slack variables ${\lambda}_{\mathcal{M}_t}$ and incorporating the following additional linear constraints
\begin{equation}
\label{main5}
 \begin{array}{*{35}{l}}
{\rm{min}}\hspace{1mm}\{0,{p}_{\mathcal{M}_t}^{\rm{lb}}\}\le {\lambda}_{\mathcal{M}_t} \le {\rm{max}} \hspace{1mm} \{0,{p}_{\mathcal{M}_t}^{\rm{up}}\}\vspace{3mm} \\
{p}_{\mathcal{M}_t}^{\rm{lb}}{c}_{\mathcal{M}_t} \le {\lambda}_{\mathcal{M}_t} \le {p}_{\mathcal{M}_t}^{\rm{ub}}{c}_{\mathcal{M}_t} \vspace{3mm} \\
{p}_{\mathcal{M}_t}-{p}_{\mathcal{M}_t}^{\rm{ub}}(1-{c}_{\mathcal{M}_t})\le {\lambda}_{\mathcal{M}_t} \le {p}_{\mathcal{M}_t}-{p}_{\mathcal{M}_t}^{\rm{lb}}(1-{c}_{\mathcal{M}_t}) 
\end{array}
\end{equation} 
Therefore, we linearize each term in the sum separately. This way we end up with ${K}{N}$ of the $\lambda$ variables, for example ${\lambda}_{\mathcal{M}_t}$.
The steps followed in the linearization of constraint C4 are:
 \begin{itemize}
\item Introducing a variable ${\lambda}_{\mathcal{M}_t}={c}_{\mathcal{M}_t}{p}_{\mathcal{M}_t}$ for each product.
\item Finding upper and lower bounds for each ${p}_{\mathcal{M}_t}$.
\item Introducing the constraints for each ${\lambda}_{\mathcal{M}_t}$ as in \eqref{main5}.
\item Substituting the ${\lambda}_{\mathcal{M}_t}$ variables into the constraint $\sum\limits_{t=1}^{{K}{L}}{c}_{\mathcal{M}_t}{p}_{\mathcal{M}_t}=P_{\rm{Max}}$.
 \end{itemize}
Therefore, due to the linearization of C4, we have these following linear constraints to be incorporated in the optimization problem.
\begin{equation}
\label{main6}
\text{C4:} \left\{ \begin{array}{*{35}{l}}
\text{}\hspace{1mm}\text{N1:} \hspace{2mm}{\lambda}_{\mathcal{M}_t}\le P_{\rm{Max}}{c}_{\mathcal{M}_t},\hspace{20mm}\forall t,\hspace{1mm}t=1,\cdots, {K}{L} \vspace{2mm} \\
\text{}\hspace{1mm}\text{N2:}\hspace{2mm}{\lambda}_{\mathcal{M}_t}\ge 0, \hspace{30.4mm}\forall t,\hspace{1mm}t=1,\cdots, {K}{L} \vspace{2mm} \\
\text{}\hspace{1mm}\text{N3:}\hspace{2mm} {\lambda}_{\mathcal{M}_t}\le {p}_{\mathcal{M}_t}, \hspace{24.5mm}\hspace{2mm}\forall t,\hspace{1mm}t=1,\cdots, {K}{L} \vspace{2mm} \\
\text{}\hspace{1mm}\text{N4:}\hspace{2mm} {\lambda}_{\mathcal{M}_t}\ge {p}_{\mathcal{M}_t}-P_{\rm{Max}}(1-{c}_{\mathcal{M}_t}),\hspace{1.0mm}\hspace{1.2mm}\forall t,\hspace{2mm}t=1,\cdots, {K}{L} \vspace{2mm} \\
\text{}\hspace{1mm}\text{N5:}\hspace{2mm} \sum\limits_{t=1}^{{K}{L}}{\lambda}_{\mathcal{M}_t}=P_{\rm{Max}}, 
\end{array}\right.
\end{equation}
Consider the first case ${c}_{\mathcal{M}_t}=0$, which means the product ${\lambda}_{\mathcal{M}_t}={c}_{\mathcal{M}_t}{p}_{\mathcal{M}_t}$ should be 0. The first pair of inequalities (N1, N2) says $0\le {\lambda}_{\mathcal{M}_t}\le 0$, forcing ${\lambda}_{\mathcal{M}_t}=0$. The second pair of inequalities (N3, N4) says ${p}_{\mathcal{M}_t}-P_{\rm{Max}}\le {\lambda}_{\mathcal{M}_t} \le {p}_{\mathcal{M}_t}$, and ${\lambda}_{\mathcal{M}_t}=0$ satisfies those inequalities. Now consider the case ${c}_{\mathcal{M}_t}=1$, so that the product should be ${\lambda}_{\mathcal{M}_t}={p}_{\mathcal{M}_t}$. The first pair of inequalities becomes $0 \le {\lambda}_{\mathcal{M}_t} \le P_{\rm{Max}}$, which is satisfied by ${\lambda}_{\mathcal{M}_t}={p}_{\mathcal{M}_t}$. The second pair says ${p}_{\mathcal{M}_t} \le {\lambda}_{\mathcal{M}_t} \le {p}_{\mathcal{M}_t}$, forcing ${\lambda}_{\mathcal{M}_t}={p}_{\mathcal{M}_t}$ as desired. This linearization approach, in particular, equates to splitting the feasible regions into two subregions, one where ${c}_{\mathcal{M}_t}=0$ and $f({c}_{\mathcal{M}_t},{p}_{\mathcal{M}_t})={c}_{\mathcal{M}_t}{p}_{\mathcal{M}_t}=0$ (trivially linear) and the other where ${c}_{\mathcal{M}_t}=1$ and $f({c}_{\mathcal{M}_t},{p}_{\mathcal{M}_t})={p}_{\mathcal{M}_t}$ (also linear).

   The geometry of the linearization process of constraint C4 is depicted in Fig.~\ref{FIG3}b. The blue circles represent the expected values of of the optimization variable ${\lambda}_{\mathcal{M}_t}$. The gray shaded areas represent the range of values that the variable can take due to linearization process. The red circles stands for the obtained values of the optimization variable. Note that the expected and obtained values of ${\lambda}_{\mathcal{M}_t}$ are perfectly matched, i.e., exact linearization is obtained. 

Now, we are left with the constraint C5 of \eqref{main3}, which is also nonlinear. Following the relationship in \eqref{notmain2}, constraint C5 can be stated as
 \begin{equation}
\prod\limits_{t=K_1{L}+1+(k-1){L}}^{K_1{L}+{L}+(k-1){L}}(1+{c}_{\mathcal{M}_t}{p}_{\mathcal{M}_t}{h}_{\mathcal{M}_t})\ge 2^{R_k^{\rm{target}}}, \hspace{5mm} k=K_1+1,\cdots, {K}
\end{equation} 
which further can be equivalently stated as the following
\begin{equation}
\label{yyy}
\text{C5: }\left(\prod\limits_{t=K_1{L}+1+(k-1){L}}^{K_1{L}+{L}+(k-1){L}}{\xi}_{\mathcal{M}_t}\right)^{1/L}\ge 2^{R_k^{\rm{target}}/L}, \hspace{5mm} k=K_1+1,\cdots, {K}
\end{equation}
where the left side of \eqref{yyy} is the geometric-mean of the optimization variables ${\xi}_{\mathcal{M}_t}$ and it is concave.
After performing linearization and/or convexification of the nonlinear constraints (outer approximating the feasible region) and the objective function, the MINLP optimization problem in \eqref{main2} now becomes a convex, which is comparatively much easier to solve.

\section{Performance Analysis}
In all simulation results presented in this section, the wireless channel is modeled as a frequency-selective channel consisting of six independent Rayleigh multipaths. The multipath components are modeled by Jakes' flat fading model \cite{Rappaport}. The power delay profile is exponentially decaying with $e^{-\alpha l}$ with $\alpha=1$(decay factor), where $l$ defines the multipath index. A maximum delay spread of 5$\mu$s and maximum doppler of 30 Hz are assumed. The relative power of the six multipath components are [0, -4.35, -8.69, -13.08, -17.43, -21.78] dB. The power spectral density of AWGN is -170 dBm/Hz. We also assume that the width of the shared spectrum is 10 MHz and it is contiguous. A common subcarrier grid is generated with 512 subcarriers and the operators employ OFDMA for carrying users' data. In the performance analysis for fragment based spectrum allocation, all the operators employ same subcarrier spacing and radio interfaces with OFDM waveforms, however, the DFT sizes vary depending on the sizes of frequency fragments (i.e., depending on the number of subcarriers as fragments are integer multiples of subcarrier bandwidth) the operators are assigned with. For example, let's say, during one scheduling, the cardinalities of the sets of subcarriers (contiguous) assigned to three spectrum sharing operators are given by $|\mathcal{S}_1|=112$, $|\mathcal{S}_2|=213$ and $|\mathcal{S}_3|=187$. Then the DFT sizes employed by operator 1, operator 2 and operator 3 are 128, 256 and 256, respectively. Note that DFT size is chosen as a power of 2 greater than or equal to the the number of subcarriers one operator is assigned with. On the other hand, in subcarrier-gain based allocation, all the operators employ DFT size of 512 irrespective of the number of subcarriers each operator is assigned with since all the operators operate over the whole shared spectrum.  Each BS may perform subchannelization of the subcarriers it is assigned with for intra-operator resource allocation if data transmission occurs in terms of resource block. Note that we compare the performance of subcarrier and fragmentation based spectrum sharing in terms of achieved throughput. Standard water-filling algorithm is followed to distribute the power optimally among the subcarriers per user per operator. The nonlinear optimization solver KNITRO \cite{KNITRO} along with mathematical programming modeling language AMPL \cite{AMPL} have been employed to solve the intra-operator resource allocation optimization problem.

\begin{figure}
  \centering
\includegraphics[scale=.08]{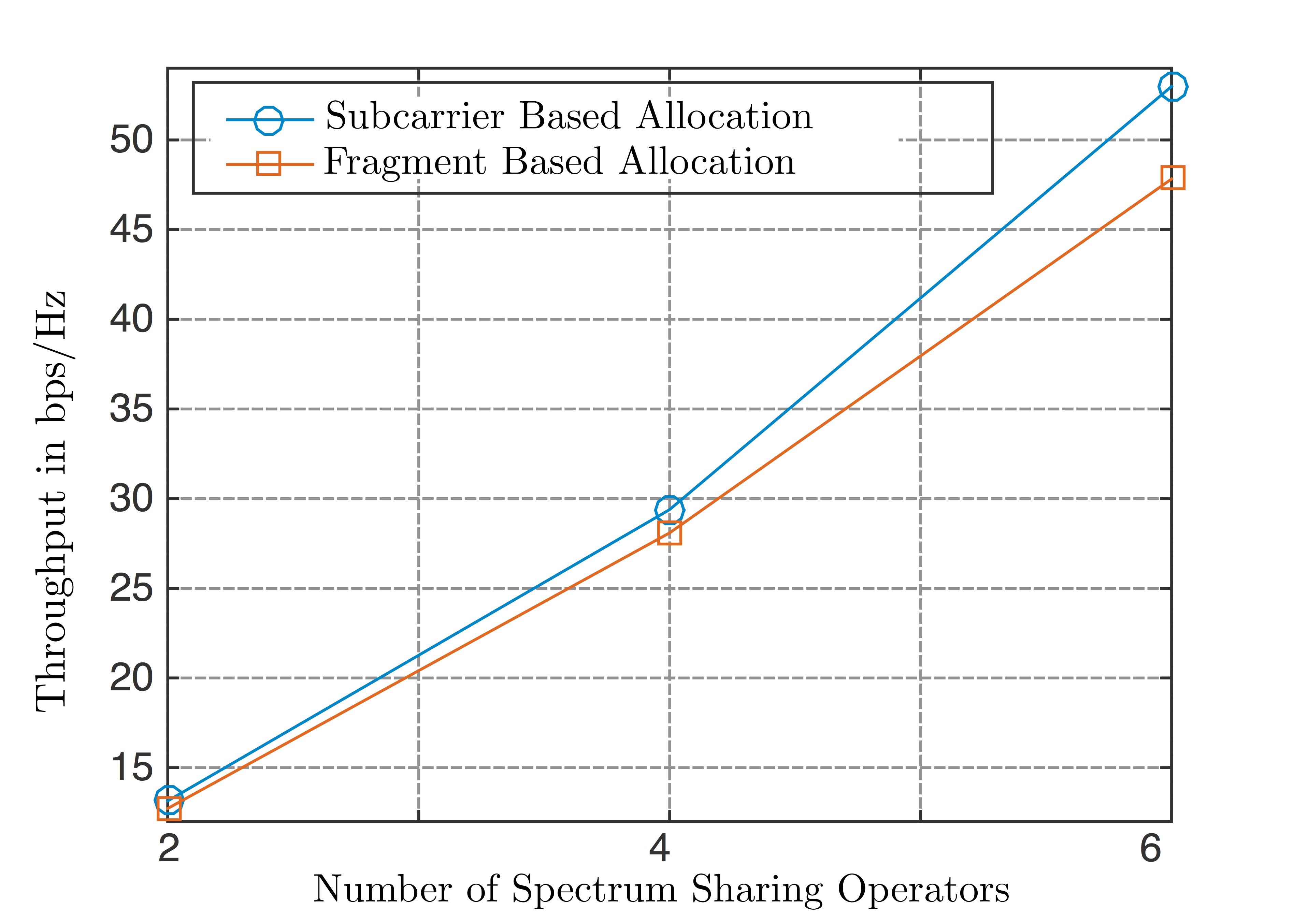}
   \caption{Impact of varying the number of spectrum sharing operators on achieved system throughput. For simplicity, we have assumed $\rho_1=\rho_2=\cdots=\rho_{N_{\rm{op}}}$ with $N_{\rm{sub}}$=512. Each operator transmits its signal with 4 watts, irrespective of the number of subcarriers it is assigned with. When there is increase in the number of operators, there is linear increase in total transit power $\sum_{n=1}^{N_{\rm{op}}}P_{\rm{Max}}^{(n)}$. For example, (i) $N_{\rm{op}}=2$, total power is 8 watts, and (ii)  $N_{\rm{op}}=6$, total transmission power in the spectrum shared access system is 24 watts, regardless of the fact that the number of subcarriers, $||\mathcal{S}_{\rm{grid}}||$ remains same.}
   \label{PR1_1}
\end{figure}

In Fig.~\ref{PR1_1}, we have investigated the impact of varying the number of spectrum sharing operators on achieved system throughput and compared the performances of subcarrier gain based and fragmentation based spectrum sharing. It can be clearly seen that subcarrier based spectrum distribution achieves higher throughput, and the performance gap increases as the number of operators is increased under a given set of subcarriers. This is due to there fact that when a large number of operators participate in the shared spectrum access communication, the multi-user diversity improves. Compared to the fragmentation based spectrum sharing, subcarrier based spectrum sharing has higher flexibility in allocating the subcarriers to the operators, i.e., to the users who have the highest gains for the subcarriers. The only disadvantageous fact about the subcarrier based distribution is that the GSC needs to send each operator the indices of the subcarriers, which, in turn, increases the backhaul load. It is worthy to mention that for the performance evaluation in Fig.~\ref{PR1_1}, we have considered a spectrum sharing scenario where all the operators are overloaded and $\rho_1=\rho_2=\cdots=\rho_{N_{\rm{op}}}$. Therefore, when there are two operators sharing the spectrum, all the available subcarriers are assigned to two operators equally. When there are more ($>2$) operators, the available spectrum is assigned to the operators in accordance with sharing rules and active priority measures. We have already mentioned that the operators have the flexibility in employing independent radio interface, e.g., multicarrier waveform, DFT size, etc. Under this circumstance, the operators employ varying DFT sizes depending on the numbers of subcarriers they are assigned with.

\begin{figure}
  \centering
   \includegraphics[scale=.18]{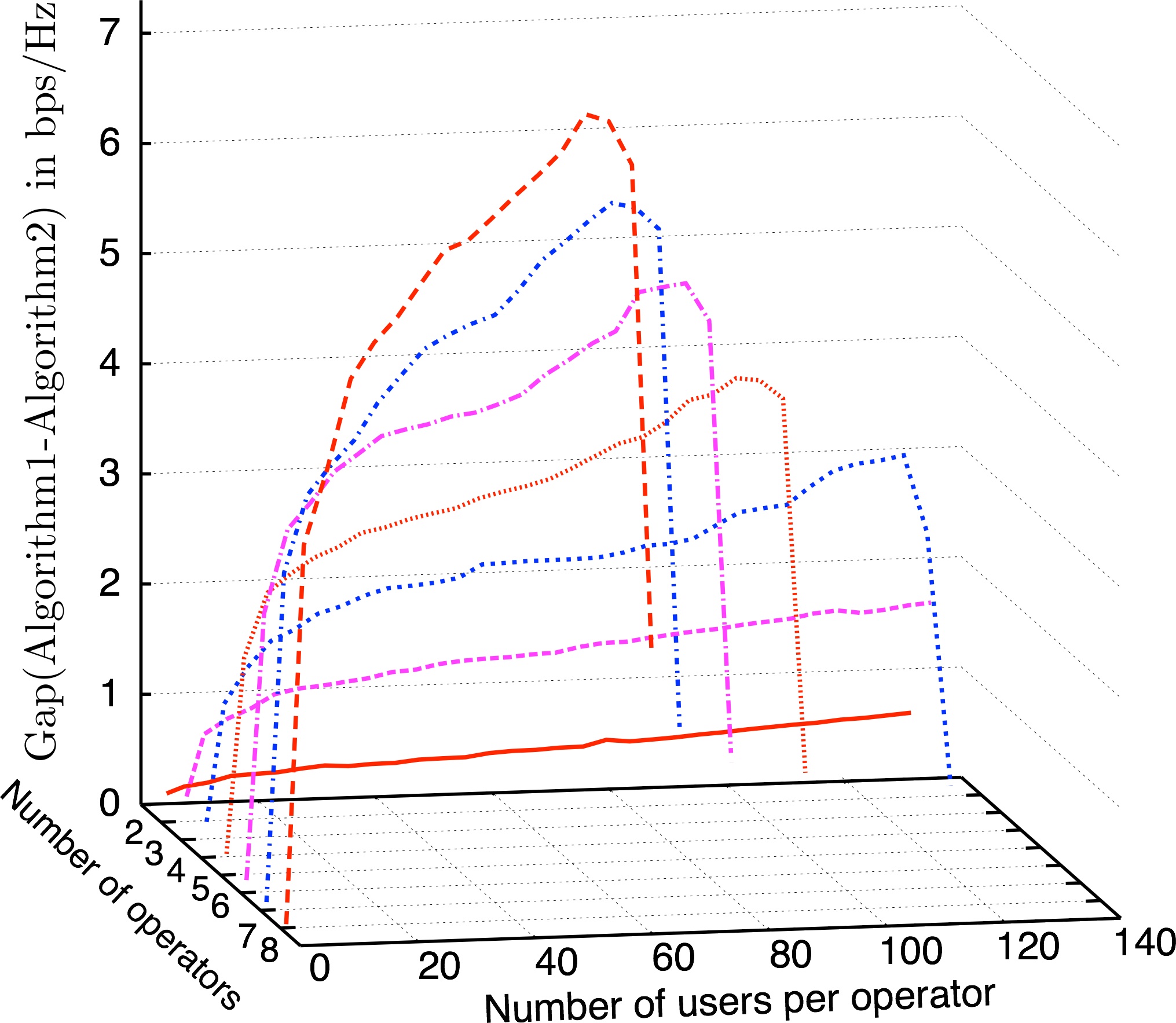}
   \caption{Multiuser diversity impact on the performances of the proposed spectrum sharing approaches. For simplicity, we have assumed $\rho_1=\rho_2=\cdots=\rho_{N_{\rm{op}}}$ with $P^{(n)}_{\rm{Max}}$=4 watts, $\forall n$. }
   \label{VaryingUsers}
\end{figure}

In Fig.~\ref{VaryingUsers}, we study the impact of multiuser diversity on spectrum shared access system in terms of difference in achieved throughputs between subcarrier gain based and fragmentation based spectrum sharing solutions by varying the number of operators and also the number of users per operator.
We can clearly distinguish three distinct regions on this figure: (i) \textit{Region-1} is determined by the steepest ascent in the performance gap curves and it corresponds to scenarios in which there are few users per operator; (ii) \textit{Region-3} is determined by the steepest descent in the performance gap curves and corresponds to scenarios in which the number of users per operator is (almost) maximum; (iii) \textit{Region-2} lies within \textit{Region-1} and \textit{Region-3} and corresponds to the scenarios most likely found in real systems.
\textit{Region-1} and \textit{Region-3} can be treated as boundary regions where both of the proposed spectrum sharing approaches perform almost equally or where the gap is small.
That is to say, \textit{Region-1} represents the, unlikely yet possible, scenario in which a single user gets assigned with all the subcarriers belonging to its operator, for both the subcarrier-based and fragment-based approaches.
On the other hand, \textit{Region-3} should be thought as the scenario in which there are as many users as the number of available subcarriers and each user is assigned with a single subcarrier.

Note the higher gap between subcarrier-based and fragment-based allocation methods is achieved when the system has more spectrum sharing operators.
This is due to the fact that the set size of available subcarriers inherent to each allocation method decreases with different rates as the number of spectrum sharing operators grows.
For instance, although there is throughput improvement due to multiuser diversity under the fragment-based solution, such improvement is significantly affected as each operator gets assigned with a smaller fragment due to the existence of more operators. 
Therefore, from the whole spectrum sharing communication system perspective, the maximum gap depends on the size of the subcarrier grid, sharing parameters $\{\rho_1,\rho_2,\cdots,\rho_{N_{\rm{op}}}\}$ and the number of spectrum sharing operators. For any particular operator, the gap will solely depend on the set size of the subcarriers it is assigned with and the number of users the operator serves.
We can still observe that the analysis shown in Fig.~\ref{VaryingUsers} could be useful for an entity managing the shared spectrum in order to make a judicious selection of allocation method. For instance, suppose that the aforementioned GSC decides that a performance gap within $1$ bps/Hz is acceptable. In a region where $3$ operators having up to 20 users each share the licensed spectrum, the fragmentation based allocation method should be utilized since the performance requirement is satisfied with reduced computational time.

\begin{figure}
  \centering
 \includegraphics[scale=.08]{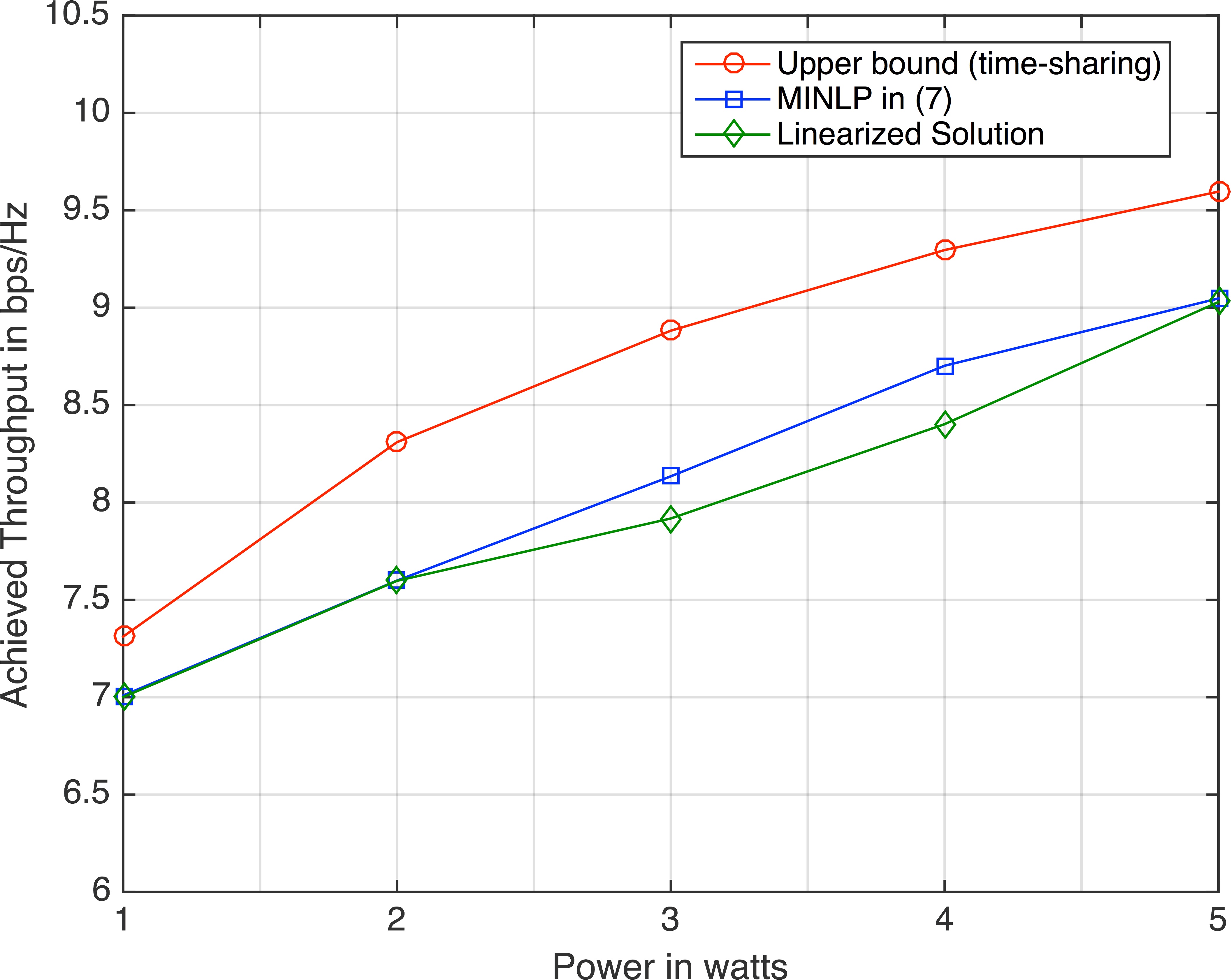}
   \caption{Comparison of spectral efficiencies achieved by solving the original MINLP and solving our proposed linearized formulation for an operator with 4 (2 DC users and 2 NDC users) users and 8 subchannels. }
   \label{PR2}
\end{figure}
We evaluate the achieved spectrum efficiency by employing the proposed linearization based intra-operator resource allocation approach for users with DC and NDC service requirements in Fig.~\ref{PR2}, and compare it with original MINLP based solution and other existing methods. Here, 2 out of total 4 users have DC service requirements of 1.4 bps/Hz and 1.6 bps/Hz, respectively.  Note that the spectrum efficiencies achieved by solving the original MINLP problem and proposed approach are very close. The upper bound curve is obtained by solving the proposed method in \cite{Tao}. The solution gives an upper bound on the achievable maximum sum-rate of all NDC users under the individual rate requirement for each DC user and the total transmit power constraint. But, the proposed convex relaxation technique in \cite{Tao} permits time-sharing of each subchannel, therefore, the system model differs from the original OFDMA system since it does not perform mutually exclusive subchannel assignment. If we strictly follow non-overlapping subcarrier allocation, the correct bound is MINLP solution. That is to say that optimal solution of the MINLP is also an optimal solution of the linearized problem, i.e., the upper bound is given by the MINLP solution. In other words, for integer programming problems (i.e., $c_{\mathcal{M}_t}\in \{0,1\}$), MINLP solution evolves as the upper bound.

Exact solution of the linearized and convex optimization problem depends on factors like tolerance of the feasibility error and relative optimality tolerance of the solver. Smaller values of absolute feasibility tolerance and relative optimality tolerance result in a higher degree of accuracy in the solution with respect to the feasibility and optimality, respectively, but the solution will be more expensive. We have kept the default values for tolerance of the feasibility error and optimality error, which is $1.0e{-6}$ for both the cases. As this value is quite small, even if we cannot guarantee the optimality, we can at least claim that the solution provided by the solver is very close to the optimal solution. 

It can be noticed that there exists a difference in performance between the original MINLP and linearized program solutions. One possible reason for this performance gap is that although in Fig.~\ref{FIG3}b we certainly obtain perfect matching of the expected and obtained values of the slack variables, we may not obtain such perfect matching in Fig.~\ref{FIG3}a. Furthermore, it should be noted that the obtained values of the slack variables in Fig.~\ref{FIG3}a are always lower than the expected/maximum values. Therefore, when the MINLP is solved, it is very likely that the expected values will be obtained. However, when the linearized problem is solved, the obtained values may not always match the expected values. Since, obtained values are always $\le$ expected values, there may exist a difference in performance of the linearized/convexified and MINLP solutions. In order to find the probable cause responsible for this gap, we followed the following concept: after optimization, if the slack variable $\xi$ is strictly smaller than the right side of (11), it would be easy to increase $\xi$ until the inequality in (11) becomes an equality. This would eventually benefit the objective function and it would not violate any constraint. The constraint C5 in (16) will even benefit from an increase of $\xi$. However, we found that the constraint in  equation (11): ${\xi}_{\mathcal{M}_t}\le 1+{\rm{min}}\{{p}_{\mathcal{M}_t},P_{\rm{Max}}{c}_{\mathcal{M}_t}\}{h}_{\mathcal{M}_t}$ is not that main cause of the performance loss. Interestingly, we noticed that the values of ${\xi}_{\mathcal{M}_t} -{\rm{min}}\{{p}_{\mathcal{M}_t},P_{\rm{Max}}{c}_{\mathcal{M}_t}\}{h}_{\mathcal{M}_t}$ are very close to 1, and even after increasing $\xi$ until the inequality in (11) becomes an equality, does not really help much in improving the gap. Thus this gap may be the consequence of numerical inaccuracies of the solver we employed.

From the point of view of computational complexity of the intra-operator resource allocation problem, as far as worst-case complexity goes, MINLP problems are provably unsolvable \cite{Jeroslow}, but linearized problems are solvable (sometimes, perhaps not very quickly). Furthermore, nonlinear programming is more difficult than linear programming, especially when the feasible region is non-convex since it can be hard to determine which points are actually optimal. Therefore, if we have a choice, a linearized formulation is probably a better choice because the constraints are easier to deal with, and the LP relaxation gives an LP optimum. However, it is not straightforward to prefer one formulation over the other just based on computational complexity. One should however make such decisions based on a judicious analysis of existing tools and algorithms, and perform experiments to find out the most efficient approach. In general, in the current state-of-the-art, the linearized formulation tends to win.

\section{Conclusions}

In this paper, we considered a shared spectrum access communication system and proposed suboptimal solutions both for inter-operator and intra-operator resource allocation problems. For inter-operator spectrum allocation, two efficient algorithms (subcarrier and fragmentation based) are proposed that take care of the mutual sharing policy and fairness issues. Subcarrier based spectrum allocation scheme has been found to be more efficient in terms of achieved throughput. However, fragmentation based allocation scheme is more suitable in terms of computational complexity. For the intra-operator resource (spectrum and power) allocation problem, we considered resource allocation for a system with users with delay constraint service requirements. and formulated computationally efficient (if not, at least solvable) solutions based on some linearization techniques (exact linearization or linear approximations) considering the structures of the optimization problems and the constraints. The performances of the proposed solutions are impressive when compared to the original MINLP and other existing solutions.

In the current status of the proposed spectrum sharing algorithms, we have not considered the loss of spectral efficiency due to additional guardband being used by the operators that employ inefficient waveforms. Using spectrally inefficient waveforms by some of the operators would decrease the overall shared spectrum access spectral efficiency and at the same time, the GSC could be unfair to the operators who employ spectrally efficient waveforms. Therefore, some mechanisms need to be developed to identify the operators and quantify the loss and thus, penalize the operators employing inefficient waveforms accordingly, which will be investigated in our future works.  

In order to simplify the user terminal complexity, the availability of large and contiguous spectrum is preferable. User terminal might require to perform aggregation of multiple frequency bands with different characteristics if large contiguous spectrum for sharing is unavailable. Furthermore, the radio frequency front end of the user terminal needs to be tuneable and configurable to operate at a particular frequency band depending on shareable spectrum availability for its operator to support different spectrum of operation. Therefore, designs of frequency agile front-end for user terminal and flexible air interface for supporting dynamic usage of spectrum, need to be investigated.

\section*{Competing interests}
  The authors declare that they have no competing interests.
  
\section*{Acknowledgment}
\small{This research was conducted under a contract of R\&D for radio resource enhancement, organized by the Ministry of Internal Affairs and Communications, Japan.}

\end{document}